\documentclass[prl,reprint,twocolumn,superscriptaddress, floatfix,showpacs]{revtex4-1}%

\usepackage{amsfonts}
\usepackage{amsmath}
\usepackage{amssymb}
\usepackage{graphicx}%

\begin{document}


\title{Model for Dynamic Self-Assembled Magnetic Surface Structures}
\author{M.~Belkin}
\affiliation{Department of Chemical Engineering,
  Northwestern University, 2145 Sheridan Rd, Evanston, IL 60208}

\affiliation{Materials Science Division, Argonne National
Laboratory, 9700 South Cass Avenue, Argonne, IL 60439}

\author{A.~Glatz}
\affiliation{Materials Science Division, Argonne National
Laboratory, 9700 South Cass Avenue, Argonne, IL 60439}

\author{A.~Snezhko}
\affiliation{Materials Science Division, Argonne National
Laboratory, 9700 South Cass Avenue, Argonne, IL 60439}

\author{I.S.~Aranson}
\affiliation{Materials Science Division, Argonne National
Laboratory, 9700 South Cass Avenue, Argonne, IL 60439}



\begin{abstract}
We propose a first-principles  model for self-assembled
magnetic surface structures on the water-air interface reported in
earlier experiments \cite{snezhko2,snezhko4}. The model is
based on the Navier-Stokes equation for liquids in shallow water
approximation coupled to Newton equations for interacting magnetic
particles suspended on the water-air interface. The model reproduces
most of the  observed phenomenology, including spontaneous formation
of magnetic snake-like structures, generation of large-scale vortex
flows, complex ferromagnetic-antiferromagnetic  ordering of the snake,
and self-propulsion of bead-snake hybrids.
The model provides valuable insights into self-organization
phenomena in a broad range of non-equilibrium magnetic and
electrostatic systems with competing interactions.
\end{abstract}

\date{\today}

\maketitle

Understanding the fundamental principles governing dynamic self-assembly
in non-equilibrium systems  continues attracting enormous attention
in physics and engineering  communities
\cite{bart,snezhko2,snezhko4,snezhko5,glotzer,monica,theor,electro}.
The interest is stimulated by the need for creating smart,
dynamic materials capable of self-assembly,  adaptation  to
environments, and for the design of artificial  structures capable of
performing useful tasks on the microscale \cite{mach1,gears},
including targeted cargo delivery~\cite{delivery1} or
stirring in microfluidic devices~\cite{microfluid}.

\begin{figure}[t]
\includegraphics[width=7cm]{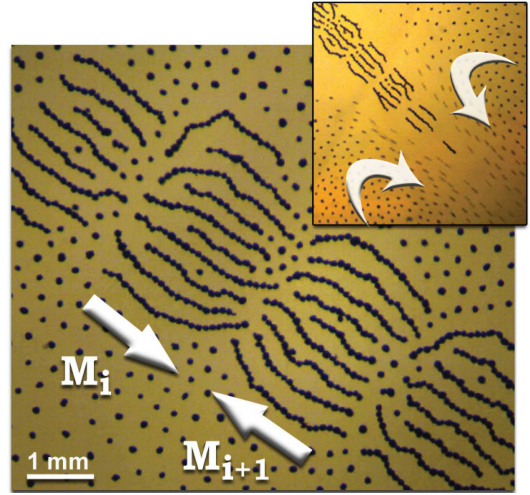}%
\caption{(Color online): Self-assembled magnetic snake generated by
an \emph{ac} field ($f=50$~Hz, $H_0=100$~Oe), composed of 90$\mu$m
Nickel spheres: individual particles form self-assembled
ferromagnetic chains which in turn align in segments of the snake.
Arrows show the orientation of the total magnetic moments in the
segments. \emph{Inset}: Pair of counter-rotating vortices at the
tail of a magnetic snake.}
\label{fig1}%
\end{figure}

In a series of works~\cite{snezhko2,snezhko4} we reported
experimental studies of self-assembled dynamic magnetic
microstructures (magnetic snakes)  formed on  the water-air
interface from a dispersion of magnetic particles energized by an
alternating (\emph{ac}) magnetic field applied perpendicular to the
interface. The snakes form due to coupling between the fluid's
surface deformations and the collective response of particles to
an \emph{ac} magnetic field. These spectacular dynamic structures
have complex magnetic ordering~\cite{snezhko6}: the snakes' segments are
formed by ferromagnetically aligned chains of microparticles;
however, the segments are always anti-ferromagnetically ordered, see
Fig.~\ref{fig1}. The snakes generate two  pairs of large-scale
vortices located at the tails. Under certain conditions snakes
spontaneously break the symmetry of the vortex pairs  and turn into
self-propelled entities~\cite{snezhko4,belkin2}. Some aspect of the
snakes' behavior were reproduced by phenomenological models based on
the Ginzburg-Landau type equation for surface waves coupled to
a large-scale flow \cite{snezhko4}. Nevertheless,  the
fundamental microscopic mechanisms leading to the formation of
snakes and the relation between properties of the snakes and
microscopic properties of the system remain unclear.

In this Letter we report on  a mathematical model capturing entire
self-assembly dynamics in a system of magnetic microparticles on the
water-air interface. The model is formulated in terms of the
Navier-Stokes equation for fluids in a shallow water approximation
coupled to interacting magnetic particles described by the
corresponding Newton equations. The particles-fluid coupling is
two-fold: an external \emph{ac} magnetic field causes oscillations
of the particles and deformations of the air-fluid interface. In
turn, hydrodynamic flows advect/reorient  particles and thus mediate
magnetic dipole interactions. Our numerical studies faithfully
reproduce observed phenomenology: formation of dynamic snake-like
arrangements of magnetic particles from initially disordered
configurations, complex ferromagnetic-antiferromagnetic ordering of
the snake, quadrupole structure of  self-generated vortex flows, and
even self-propulsion of a bead-snake hybrid. The computational
algorithm is implemented for graphics processing units (GPUs).
Insight and techniques developed in the course of this work can be
applied to a variety of non-equilibrium interfacial non-equilibrium
systems with competing  interactions.
\begin{figure}[t]
\includegraphics[width=7cm]{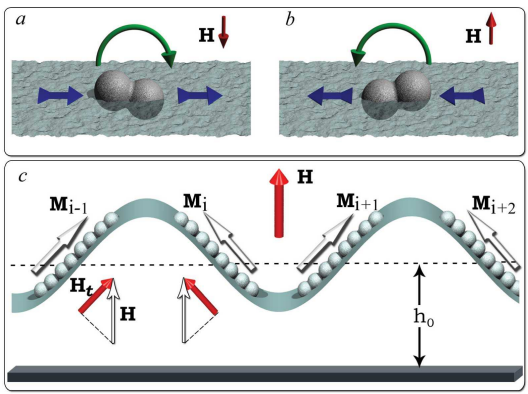}%
\caption{(Color online): {(a), (b)}: Surface flows
generated by dimers/short chains  due to  external torques exerted
by a vertical \emph{ac} magnetic field. {(c)}: coupling between
chains and  surface deformations. White arrows indicate the magnetic
moment orientations in segments. \textbf{$H_ t$} is a tangential
component of the external \emph{ac} field enforcing a specific orientation of
the chain's magnetic moment.
}
\label{fig2}%
\end{figure}

A precise  description of the motion of particles suspended on the
air-water interface is computationally prohibitive. In particular,
accurate modeling of the three-dimensional finite Reynolds number (Re)
Navier-Stokes equation (snakes have Re $\approx100$), realistic
account for particle-fluid interaction and description of
solid-fluid contact lines required enormous computational power. Here we
propose a simplified yet nontrivial model allowing for the
investigation of the entire process of the snakes' self-assembly.
The model identifies key physical mechanisms and
ingredients. We made the following major approximations: (i) we
considered the Navier-Stokes equation in the so-called shallow water limit (see, e.g.
\cite{landau}).
This approximation is valid when the
characteristic length of the snake is large compared to the thickness
of the fluid layer $h_0$. We verified independently  in our experiments
that the snakes exist on thin layers of liquid.
(ii) We simplified
the magnetic interaction between  particles by
 restricting orientation of their magnetic moments along the surface of the fluid.
 This implies that  the ``elementary'' object in our
description is  a microchain or a magnetic dimer formed by
two spherical ferromagnetic particles (the magnetic moment of the chain is
 directed along the chain).
(iii) We  simplified the particle-fluid interaction by assuming that
the  primary effect of the vertical \emph{ac}  magnetic  field is a
rocking motion of the particles leading to the generation of
surface flows, see Fig.~\ref{fig2}a,b. We took into
account advection of particles by surface flows and rotation of
their orientation due to vorticity and shear. However, we neglected
hydrodynamic flows due to in-plane drift and rotation of the
particles: for the conditions of our experiments, the magnitude of
these motions is substantially smaller than those produced by the
external magnetic field.

The evolution of the fluid surface $h$ and the
depth-averaged hydrodynamic velocity $v$ is described by
\begin{eqnarray}
\partial_t h + \nabla(h{\bf v}) &=& 0,\label{shallow_water_h}\\
\partial_t{\bf v} + ({\bf v}\nabla){\bf v} &=&
\eta(\nabla^2{\bf v} - \xi{\bf v}) - \nabla h + \sigma\nabla\Delta h  \nonumber \\
&& + H_0\sin(\omega t)\sum_j s({\bf r} - {\bf r}_j){\bf p}_j\label{shallow_water_v}
\end{eqnarray}
where $\eta$ is the kinematic  viscosity; $\sigma$  the surface tension;
the term $-\xi{\bf v}$ describes  friction with the bottom
of a container (we used $\xi=3$ for laminar flow). Eq.
(\ref{shallow_water_v}) is linearized near the equilibrium height
$h_{0}$. The variables are scaled as follows: coordinates ${\bf r}  \to
{\bf r } /h_{0}$, time $t \to t \sqrt{ h_{0}/g}$, velocities ${\bf v}
\to {\bf v }/\sqrt {g h_{0}} $, where $g$ is the gravitational
acceleration. The last term in Eq. (\ref{shallow_water_v}) describes the
forcing on the fluid induced by motion of particles in the applied
\emph{ac} magnetic field with the magnitude $H_0$ and frequency
$\omega$. Here, ${\bf p}=(\cos(\phi), \sin(\phi)) $ is a unit
orientation vector parallel to the particle's  magnetic moment  and
$s(r)$ is a  function describing  the shape of the
particle. In most of our simulations we used: $s =A_0 \exp (-r^2
/s_0^2)$ for $r<r_0 $ and $s=0$ for $r>r_0$, where $s_0 $ is the
parameter related to the particle size, $r_0$ is a cutoff radius,
typically 5-7 particles sizes, and the parameter $A_0$ (set later to $A_0=1$)  characterizes
the  strength of the magnetic forcing. Our simulations show that this
particular choice of the  function  $s$ provides a good coupling
between particles and the continuum hydrodynamic field.

\begin{figure*}
\includegraphics[width=13 cm]{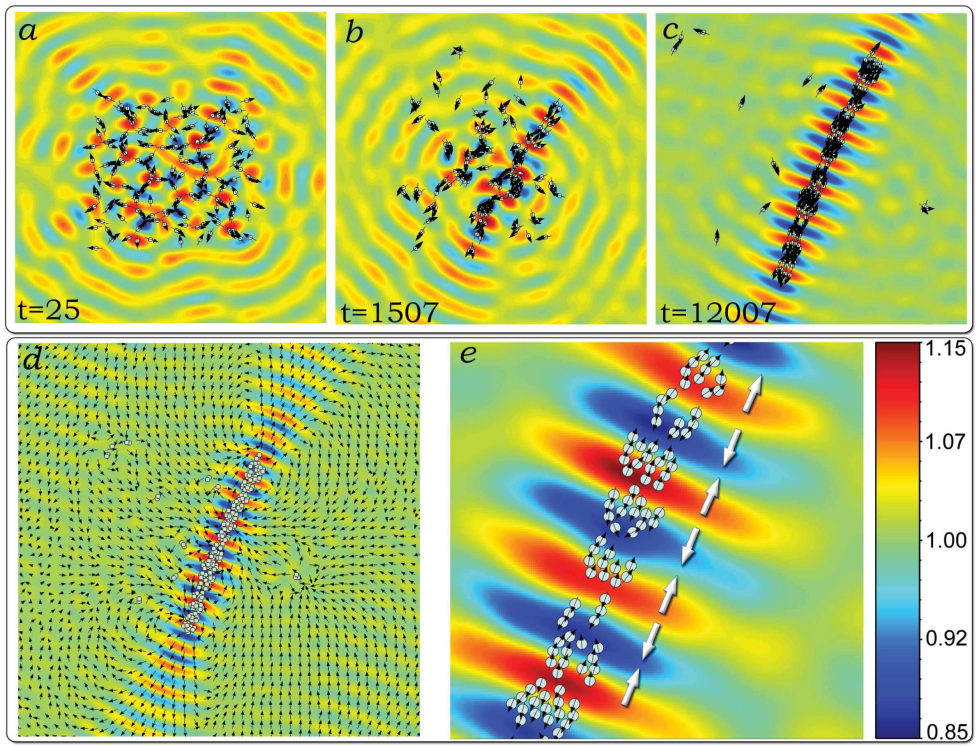}
\caption{(Color online): (a)-(c) Formation of a
snake from 225 initially randomly distributed particles. The
background color represents the height of the fluid surface
~(\emph{h}),  black arrows depict the orientation of the magnetic
moments of the particles, and particles are shown as grey circles.
Only a part of the entire integration domain is shown. (d) Flows generated by the snake in the entire domain. (e)
Antiferromagnetic order between the snake's segments; chains of
particles are ferrromagnetically ordered in each segment. The
parameters of the simulations are:
 amplitude $H_0=0.56$, frequency $\omega=1$, layers thickness $h_0=1$, domain area
$160 \times 160$, magnetic moment $\mu_0=0.26$, viscosity $\eta=0.01$, and particle diameter $a=0.8$, and $\kappa=3$ (Movies and other parameters are in \cite{online}).
These parameters are estimated for the following experimental values: layer depth 1 mm, particles radius 0.4 mm, frequency of the ac field $f \approx 18 $ Hz ($\omega=2 \pi f$), and for the viscosity and surface tension of water.}
\label{fig3}%
\end{figure*}

Particle positions and orientations  ${\bf r}_j, {\bf
p}_j$ are governed by  the following equations \cite{pedley1992}
\begin{eqnarray}
m_p\ddot{\bf r}_j + \mu_t \dot{\bf r}_j& =& {\bf F}_{j} +\mu_t {\bf v} - \beta\nabla h \label{eq_v} \\
I_p\ddot{\phi}_j + \mu_r\dot{\phi}_j& =& T_{j} + \frac{\mu_r \Omega}{2} +\nonumber \\
 \mu_r \epsilon {\bf p}_j \cdot {\bf E} \cdot ({\bf I} && - {\bf
p}_j{\bf p}_j)\times {\bf p} _j  +
 \kappa H_0\sin(\omega t)\nabla
h\times{\bf p}_j \;\; \; \; \label{eq_phi}
\end{eqnarray}
where  $m_p$, $I_p$, $\mu_t$, and $\mu_r$ are the particle's  mass,
moment of inertia, translational  and  rotational viscous drag
coefficients, respectively. ${\bf F}_{j}$ and $T_{j}$ are forces and
torques due to magnetic dipole-dipole interaction and steric
repulsion between the particles (torques have only one non-zero
component). $\Omega=\partial v_y/\partial x - \partial v_x/\partial
y$ is the vorticity and $E_{kl} = (\partial v_k/\partial x_l +
\partial v_l/\partial x_k)/2 $ is the rate of the strain tensor of
the hydrodynamic flow, and $\bf I$ is the identity matrix. While we
kept --  for the sake of completeness -- the accelerations
$m_p\ddot{r}_j,I_p\ddot{\phi}_j$ in the equations of motion, for
typical experimental conditions these terms are irrelevant and can
be omitted. Eqs.~(\ref{eq_v})\&(\ref{eq_phi}) have the following
meaning: in addition to magnetic forces and torques ${\bf F}_{j},
T_{j}$, the particles are subject to advection by the hydrodynamic
flow ($\sim {\bf v}$), sliding down the gradient of the surface due
to gravity ($\beta \nabla h$) and rotation due to the flow's
vorticity ($\sim \Omega$). Moreover, anisotropic objects, such as
microchains formed by magnetic particles, will tend to align along
the axis of the rate of the strain tensor $E_{kl}$ with a certain
prefactor  $\epsilon$ which in turn depends on the shape of the
particle (for spheres $\epsilon=0$). The last term in
Eq.~(\ref{eq_phi}) deserves a special attention: it describes the
magnetic alignment of particles along the direction of projection of
the external \emph{ac} magnetic field {\it parallel} to the surface of the
fluid, see Fig.~\ref{fig2}c. The in-plane component of the field,
oscillating with the same frequency $\omega$ appears due to the
deformation of the surface, and is proportional to $\nabla h$. This
in-plane field promotes {\it antiferromagnetic} ordering of the
neighboring segments and is crucial for the formation of snakes.

The interaction between particles $j,k$  is described by the
Hamiltonian ${\cal H} = {\cal H}^d + {\cal H}^h$, where ${\cal H}^d
$ is due to the magnetic dipole-dipole interaction,
\begin{equation}
{\cal H} ^d  = -\frac{\mu_0^2 }{4\pi r^3_{jk}}\left[3({\bf p}_j{\bf e}_{jk})({\bf p}_k{\bf e}_{jk}) - {\bf p}_j{\bf p}_k\right],
\label{magn}
\end{equation}
$r_{jk}$ is distance between particles,
 ${\bf e}_{jk}={\bf r}_{jk}/r_{jk}$, $\mu_0$ magnetic moment of individual particle, and
 ${\cal H}^h = \mu_0^2 (a/r_{jk})^{24}/16 \pi a^3 $ models a sufficiently rigid short-range hard-core repulsion between the
particles of diameter $a$ \cite{Hucht2007}. Correspondingly, forces
are evaluated as ${\bf F}_j  = - \partial {\cal H} /\partial  {{\bf
r }_j} $ and torques as $T_j = -
 {\bf p}_j \times   \partial {\cal H} /\partial  {{\bf p }_j} $.

Equations (\ref{shallow_water_h})-(\ref{eq_phi}) were solved in a
periodic $x,y$ domain by a quasispectral method. We
used a domain area of $160^2$ in dimensionless units (the length is normalized by the layer height $h_{0}$), on a grid with $1024^2$ points, and up to 225  magnetic
particles. The algorithm was implemented for massive parallel GPUs and run on a NVIDIA  GTX285
GPU with a peak performance of  1 TFlop.
Typically, a speed up
more than 100 times was achieved compared to a fast Intel i7 CPU, for details on the implementation of the algorithm see
\cite{online}.

Figures~\ref{fig3}(a)-(c) illustrate the formation of a snake from
initially random  distributed particles. The result qualitatively
reproduces experimental observations: magnetic particles assemble
into short chains, then chains form segments consisting of
several parallel chains, and finally segments form snakes-like linear
objects. Like in experiments, the segments  are
ordered {\it anti-ferromagnetically} Fig.~\ref{fig3}~(e).  We also
plotted the mass flux vector $h {\bf v}$ (unlike in an experiment,
Eq.~(\ref{eq_v}) does not provide information on the surface velocity).
Remarkably, we observed four large vortices at the tails of the
snake, see Fig.~\ref{fig3}~(d).

\begin{figure}[t]
\includegraphics[width=7 cm]{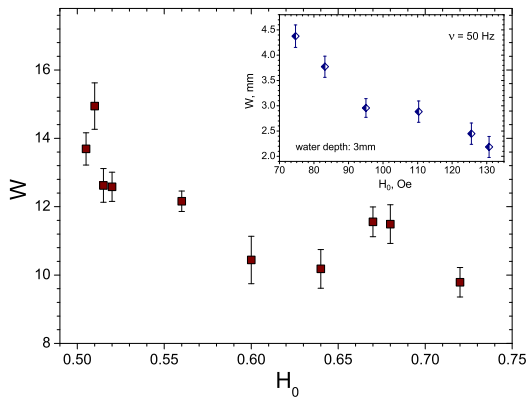}%
\caption{Snake's width  $W$  vs magnetic field
amplitude $H_{0}$ in simulations (main plot) and in experiment
(inset).}
\label{fig4}%
\end{figure}

We investigated the effect of the amplitude $H_0$ and frequency
$\omega$ of the external  magnetic field on the snake structure.
Good qualitative  agreement was obtained: like in experiments (the
inset of Fig.~\ref{fig4}), we observed a reduction of the snake's
width with the increase of  $H_0$, see Fig.~\ref{fig4}. An increase
of the amplitude $H_0$ above a certain threshold resulted in the
breakdown of a single snake and formation of multiple snakes with
some average length which further decreased with the increase of
$H_0$, also in agreement with experiments. An increase of the
frequency $\omega$ resulted in a decrease of the segment's size in
accordance with the dispersion relation. We also studied the effects
of variation of the particle's magnetic moment $\mu_0$. We noticed
that the snakes are formed only in a certain range of $\mu_0$. For
small $\mu_0$ no formation of chains was observed, and for too large
$\mu_0$  the magnetic forces overwhelmed hydrodynamic interactions,
and the dominant structures were long chains and closed
rings~\cite{snezhko5}.

\begin{figure}[t]
\includegraphics[width=8.5 cm]{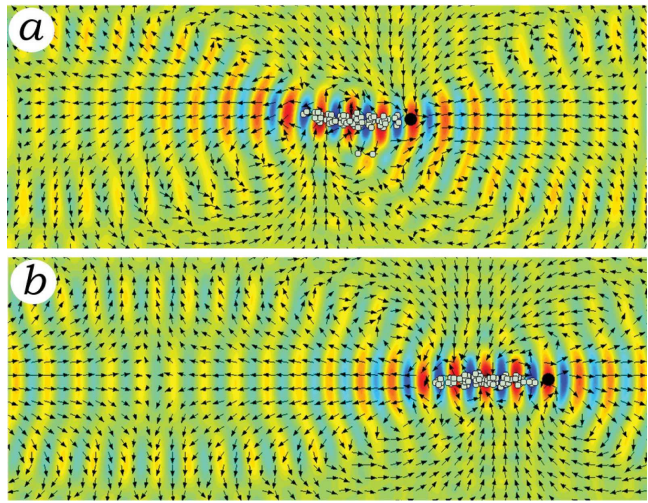}%
\caption{(Color online): Snake-bead hybrid; the
non-magnetic bead is shown as large black circle, the snake consists of 64 particles and  $H_0=0.65$ (other parameters as in
Fig. \ref{fig4}), see also Movies 2,3 in \cite{online}. The time
difference between frames is 85250. (a) and (b) show the same area of the simulated system.}
\label{fig5}%
\end{figure}

Our experiment \cite{snezhko4} revealed swimmers formed by a snake
with a non-magnetic bead. The bead at one of the snake's tails
breaks the balance between vortex flows and turns the snake into a
self-propelled entity.   To verify that, we modified one of the
particles in our simulations: the last term $H_0 \sin (\omega t) $
in Eq.~(\ref{shallow_water_v}) for  the last particle was replaced
by $-\eta \zeta_1  v({\bf r} _j)$ (i.e. an increased friction around
this particle with a certain friction coefficient $\zeta_1$); also the size of
this particle was increased by  a factor of 6 compared to magnetic
particles. The effect of self-propulsion of a snake with a bead was
successfully captured by our model, see Fig.~\ref{fig5}. Remarkably,
a swimming  snake-bead hybrid was formed from random initial
conditions: the non-magnetic bead is expelled to the periphery,
spontaneously attaches to  one of the tails of the snake and forms a
swimmer (note asymmetry between the fore and aft vortex pairs).

In conclusion, we developed a microscopic  model for self-assembled
dynamic magnetic structures at the water-air interface. We
identified the minimal ingredients necessary for the description  of
a highly nontrivial process of dynamic self-assembly. Our work
reveals that the snakes are formed due to a subtle balance between
magnetic and hydrodynamic forces. The
concepts can be applied to a wide range of interfacial particle
systems driven far from equilibrium by external forces, both on the
micro and nano scales. The research was supported by  the U.S.
Department of Energy, Office of Basic Energy Sciences, Division of
Materials Science and Engineering, under the Contract No. DE
AC02-06CH11357.

\end{document}